\documentclass[12pt]{article}
\usepackage{latexsym}
\newcommand{\be}{\begin{equation}}
\newcommand{\ee}{\end{equation}}
\def\n{\noindent}
\catcode `\@=11
\catcode `\@=12
\begin{document}
\begin{center}
\large{\bf {CONFORMALLY FLAT SPHERICALLY SYMMETRIC 
           COSMOLOGICAL MODELS-REVISITED}} \\
\vspace{10mm}
\normalsize{ANIRUDH PRADHAN $^1$ AND OM PRAKASH PANDEY} \\
\normalsize{\it{Department of Mathematics, Hindu Post-graduate College 
(V. B. S. Purvanchal University), Zamania-232 331, Ghazipur, India.}} \\
\normalsize{$^1$\it{E-mail : acpradhan@yahoo.com, pradhan@iucaa.ernet.in}}\\
\end{center}
\vspace{10mm}
\begin{abstract} 
Conformally flat spherically symmetric cosmological models representing 
a charged perfect fluid as well as a bulk viscous fluid distribution have 
been obtained. The cosmological constant $\Lambda$ is found positive and is
a decreasing function of time which is consistent with the recent supernovae 
observations. The physical and geometrical properties of the models are  
discussed.
\end{abstract}
\smallskip
\n Key Words : Cosmology; Conformally Flat Universe; Spherically Symmetric Models.\\
\n PACS No. : 98.80
\newpage
\section{Introduction}
A considerable interest has been shown to the study of physical properties of 
spacetimes which are conformal to certain well known gravitational fields. The 
general theory of relativity is believed by a number of unknown functions - the 
ten components of $g^{ij}$. Hence there is a little hope of finding physically 
interesting results without making reduction in their number. In conformally 
flat spacetime the number of unknown functions is reduced to one. The 
conformally flat metrices are of particular interest in view of their 
degeneracy in the contex of Petrov classification. A number of conformally  
flat physically significant spacetimes are known like Schwarzschild interior 
solution and Lema\^{i}tre cosmological universe.\\

At the present state of evolution, the universe is spherically 
symmetric and the matter distribution in it is isotropic and homogeneous. 
Buchdahl [1] has obtained the conformal flatness of the Schwarzschild 
interior solution. Singh and Roy [2] have discussed the possibility of 
existence of electromagnetic fields conformal to some empty spacetime. Singh  
and Abdussattar [3] have obtained a non-static generalization of 
Schwarzschild interior solution which is conformal to flat spacetime. 
Roy and Bali [4] have obtained the solution of Einstein's 
field equations representing non-static spherically symmetric perfect fluid  
distribution which is conformally flat. Pandey and Tiwari [5] have 
discussed conformally flat spherically symmetric charged perfect fluid  
distribution. Reddy [6] and Rao and Reddy [7] discussed static 
conformally flat solutions in the Brans-Dicke and Nordtvedt-Barker scalar-tensor  
theories. Shanthi [8] has shown that the most general conformally flat 
static vacuum solution in the Nordtvedt-Barker scalar-tensor theory is simply  
the empty flat spacetime of general relativity. There has been a recent literature
(Melfo and Rago [9], Mannhelm [10], Yadav and Prasad [11], Endean [12,13], 
Obukhov et al. [14], Mark and Harko [15]) which shows a significant interest in 
the study of conformally flat spacetime.\\ 

Most cosmological models assume that the matter in the universe can be described 
by `dust'(a pressureless distribution) or at best a perfect fluid. Nevertheless,
there is good reason to believe that - at least at the early stages of the universe 
- viscous effects do play a role [17]-[19]. For example, the 
existence of the bulk viscosity is equivalent to slow process of restoring equilibrium 
states [20]. The observed physical phenomena such as the large entropy per 
baryon and remarkable degree of isotropy of the cosmic microwave background radiation 
suggest analysis of dissipative effects in cosmology. Bulk viscosity is associated with 
the GUT phase transition and string creation. Thus, we should consider the presence of
a material distribution other than a perfect fluid to have realistic cosmological models
(see Gr\o n [32]) for a review on cosmological models with bulk viscosity). The effect 
of bulk viscosity on the cosmological evolution has been investigated by a number of 
authors in the framework of general theory of relativity [21]-[31].\\  

The purpose of this paper is to apply conformally flat spherically symmetric line element 
to charged perfect fluid and to bulk viscous fluid models in cosmology. This paper is 
organized as follows. The field equations are presented in Section $2$. Section $3$ includes 
the solution of the field equations in presence of charged perfect fluid distribution. 
Section $3.1$ contains some physical properties of the model. Finally in Section $4$ 
the bulk viscous models are considered.\\ 
 
\section{Field  Equations}
 We consider the conformal metric in spherical polar coordinates \\  
\begin{equation} 
\label{eq1}  
ds^{2} = e^{\lambda}(dr^{2} + r^{2}d\theta^{2} + r^{2} sin^{2}\theta d\phi^{2} -dt^{2}), 
\end{equation} 
where $\lambda$ is a function of $r$ and $t$. We number the coordinates as 
$x^{1} = r$, $x^{2} = \theta$, $x^{3} = \phi$ and $x^{4} = t$.\\  
   The energy momentum tensor for distribution of a charged perfect fluid has the form  
\begin{equation} 
\label{eq2}      
T_{ij} =  (\epsilon + p) v_{i} v_{j} + p g_{ij} + E_{ij}, 
\end{equation}  
where $E_{ij}$ is the electromagnetic field given by  
\begin{equation} 
\label{eq3}      
E_{ij} = \frac{1}{4\pi}\left[F_{ai} F_{bj} g^{ab} - \frac{1}{2} g_{ij} F_{ab}F^{ab}\right]. 
\end{equation}  
Here $\epsilon$ and p are the energy density and isotropic pressure respectively  
and $v^{i}$ is the flow vector satisfying the relation  
\begin{equation} 
\label{eq4} 
g_{ij} v^{i}v^{j} = - 1. 
\end{equation}  
The electromagnetic field tensor $F_{ij}$ satisfies Maxwell's equations  
\begin{equation} 
\label{eq5} 
F^{ij}_{;j} = 4\pi\rho v^{i}, 
\end{equation}  
\begin{equation} 
\label{eq6} 
F_{[ij;k]} = 0, 
\end{equation} 
$\rho$ being the current density. Here and henceforth a comma and a semicolon   
denote ordinary and covariant differentiation respectively. The Einstein field  
equations 
\begin{equation} 
\label{eq7} 
R_{ij} - \frac{1}{2}g_{ij}R + \Lambda g_{ij} = - 8\pi T_{ij}, 
\end{equation}  
for the line element (1) has been set up as  
\[
8\pi[(\epsilon + p)v^{2}_{1} + p e^{\lambda}] = \frac{3\lambda^{2}_{1}}{4}  
+ \frac{2\lambda_{1}}{r} 
- \lambda_{44} - \frac{\lambda^{2}_{4}}{4} \]
\begin{equation} 
\label{eq8} 
+ e^{-\lambda}(F_{14})^{2} + \Lambda e^{\lambda}, 
\end{equation} 
\begin{equation} 
\label{eq9} 
8\pi pe^{\lambda} = \lambda_{11} + \frac{\lambda^{2}_{1}}{4} + \frac{\lambda_{1}}{r} 
- \lambda_{44} - \frac{\lambda^{2}_{4}}{4} - e^{-\lambda}(F_{14})^{2} + \Lambda e^{\lambda}, 
\end{equation} 
 \[
8\pi[(\epsilon + p)v^{2}_{4} - p e^{\lambda}] = \frac{3\lambda^{2}_{4}}{4} - \lambda_{11} -   
\frac{\lambda^{2}_{1}}{4}- \frac{2\lambda_{1}}{r} \] 
\begin{equation} 
\label{eq10}
-  e^{-\lambda}(F_{14})^{2}  
- \Lambda e^{\lambda}, 
\end{equation} 
\begin{equation} 
\label{eq11} 
8\pi(\epsilon + p)v_{1}v_{4} = \frac{\lambda_{1}\lambda_{4}}{2} - \lambda_{14}. 
\end{equation} 
Equation (\ref{eq4}) gives  
\begin{equation} 
\label{eq12} 
v^{2}_{4} - v^{2}_{1} = e^{\lambda}. 
\end{equation} 

\section{Solutions of the field equations}
\noindent From eqs. (\ref{eq8}) and (\ref{eq9}) we have
\begin{equation}
\label{eq13}
8\pi[(\epsilon + p)v^{2}_{1}] - 2e^{-\lambda}(F_{14})^{2} = \frac{\lambda^{2}_{1}}{2} 
+ \frac{\lambda_{1}}{r} - \lambda_{11}.
\end{equation}
Also eqs. (\ref{eq9}) and (\ref{eq10}) readily give 
\begin{equation}
\label{eq14}
8\pi[(\epsilon + p)v^{2}_{4}] + 2e^{-\lambda}(F_{14})^{2} = \frac{\lambda^{2}_{4}}{2} 
- \frac{\lambda_{1}}{r} - \lambda_{44}.
\end{equation}
Combining  eqs. (\ref{eq12}), (\ref{eq13}) and (\ref{eq14}) we obtain
\[
8\pi[(\epsilon + p)e^{\lambda}] + 4e^{-\lambda}(F_{14})^{2} = \frac{\lambda^{2}_{4}}{2} 
- \frac{\lambda^{2}_{1}}{2} - \frac{2\lambda_{1}}{r} \] 
\begin{equation}
\label{eq15}
- \lambda_{44} + \lambda_{11}.
\end{equation}
Equations (\ref{eq9}) and (\ref{eq15}) together reduce to
\begin{equation}
\label{eq16}
8\pi \epsilon e^{\lambda} + 3e^{-\lambda}(F_{14})^{2} = \frac{3}{4}\left(\lambda^{2}_{4} 
- \lambda^{2}_{1} - \frac{4\lambda_{1}}{r}\right).
\end{equation}
In comoving coordinate system $v_{1} = 0$, then eq. (\ref{eq13}) reduces to
\begin{equation}
\label{eq17}
-e^{-\lambda}(F_{14})^{2} = \frac{\lambda^{2}_{1}}{4} + \frac{\lambda_{1}}{2r} - 
\frac{\lambda_{11}}{2}.
\end{equation}
From eq.(\ref{eq11}) we obtain
\begin{equation}
\label{eq18}
2\lambda_{14} - \lambda_{1}\lambda_{4} = 0.
\end{equation}
The general solution of (18) is obtained as
\begin{equation}
\label{eq19}
e^{\lambda} = [\alpha(r) + \beta(t)]^{-2},
\end{equation}
where $\alpha$ and $\beta$ are functions of $r$ and $t$ respectively.\\
\noindent Hence the geometry of the spacetime (1) reduces to the form
\begin{equation}
\label{eq20}
ds^{2} = \frac{1}{(\alpha + \beta)^{2}}\left(dr^{2} + r^{2}d\theta^{2} + 
r^{2}sin^{2}\theta d\phi^{2} - dt^{2}\right),
\end{equation}
which is the model for a distribution of charged perfect fluid with the flow vector
in $t$-direction. The pressure and density for the model (20) are given by
\begin{equation}
\label{eq21}
8\pi p - \Lambda = 3(\alpha^{2}_{1} - \beta^{2}_{4}) + (\alpha + \beta)
\left(2\beta_{44} - \alpha_{11} - \frac{3\alpha_{1}}{r}\right),
\end{equation}
\begin{equation}
\label{eq22}
8\pi \epsilon  + \Lambda = 3(\beta^{2}_{4} - \alpha^{2}_{1}) + 3(\alpha + \beta)
\left(\alpha_{11} + \frac{\alpha_{1}}{r}\right).
\end{equation}
Let us assume that the fluid obeys an equation of state of the form
\begin{equation}
\label{eq23}
p = \gamma \epsilon,
\end{equation}
where $\gamma(0 \leq \gamma \leq 1)$ is constant. Using eq. (\ref{eq23}) in eqs. (\ref{eq21}) 
and eq. (\ref{eq22}), we get 
\begin{equation}
\label{eq24}
\epsilon = \frac{(\alpha + \beta)}{8\pi (1 + \gamma)} (\beta_{44} + \alpha_{11}),
\end{equation}
\[
\Lambda = -\frac{(1 - \gamma)}{(1 + \gamma)} (\alpha + \beta) (\beta_{44} + \alpha_{11})
- 3(\alpha^{2}_{1} - \beta^{2}_{4}) \]
\begin{equation}
\label{eq25}
 + (\alpha + \beta)(\beta_{44} - 2\alpha_{11}).
\end{equation}
\noindent If we put $\Lambda = 0$ in our solution, we recover the solution obtained 
by Pandey and Tiwari\cite{ref5}.\\
{\bf Particular cases:}\\
{\bf Case (i):}
\noindent If we consider  $\beta(t) = \frac{a}{t^{2}}; 
\alpha, a > 0$, eqs. (\ref{eq24}) and (\ref{eq25}) reduce to
\begin{equation}
\label{eq26}
\epsilon = k_{1} \alpha t^{-4} + k_{1} a t^{-6},
\end{equation}
\begin{equation}
\label{eq27}
\Lambda = k_{2} \alpha t^{-4} + a(k_{2} + 12a)t^{-6},
\end{equation}
where $k_{1} = \frac{3a}{4\pi(1 + \gamma)}, k_{2} = \frac{12a^{2}\gamma}{(1 + \gamma)}$.\\
{\bf Case (ii):}
\noindent If we consider  $\beta(t) = \frac{a}{t}; 
\alpha, a > 0$, eqs. (\ref{eq24}) and (\ref{eq25}) reduce to
\begin{equation}
\label{eq28}
\epsilon = \alpha k_{3} t^{-3} + ak_{3}t^{-4},
\end{equation}
\begin{equation}
\label{eq29}
\Lambda = \alpha k_{3}t^{-3} + a(3a + k_{4})t^{-4},
\end{equation}
where $k_{3} = \frac{a}{4\pi(1 + \gamma)}, k_{4} = \frac{2a\gamma}{(1 + \gamma)}$.
It is observed from eqs. (\ref{eq27}) and (\ref{eq29}) that the cosmological 
constant $\Lambda$, in both cases, is a decaying function of 
time and it approaches a small value as time progresses (i.e., the present epoch),
which explains the small value of $\Lambda$ at present. Additionally, $\Lambda$
also comes out positive which is consistent with the recent supernovae observations
(Perlmutter et al. [33], Riess et al. [34], Garnavich et al. [35], Schmidt et al.
[36]).
\subsection {Physical properties of the model}
\noindent The non-vanishing component of the flow vector, $v^{4}$ is given by
\begin{equation}
\label{eq30} 
v_{4} = \frac{1}{(\alpha + \beta)}.
\end{equation}
The reality conditions $(\epsilon + p) > 0$ and $(\epsilon + 3p) > 0$ lead to
\begin{equation}
\label{eq31}
\beta_{44} + \alpha_{11} + \frac{\alpha_{1}}{r} > 0,
\end{equation}
and
\begin{equation}
\label{eq32}
(\alpha^{2}_{1} - \beta^{2}_{4}) + (\alpha + \beta) \left(\beta_{44} - 
\frac{\alpha_{1}}{r}\right) > 0.
\end{equation}
Using eq. (\ref{eq19}) in eq. (\ref{eq17}) gives
\begin{equation}
\label{eq33}
F_{14} = \frac{\left(\frac{\alpha_{1}}{r} - \alpha_{11}\right)^
{\frac{1}{2}}}{(\alpha + \beta)^{\frac{3}{2}}}.
\end{equation}
From eqs. (\ref{eq5}) and (\ref{eq33}) the current density $\rho$ is given by 
\begin{equation}
\label{eq34}
\rho = -\frac{(\alpha + \beta)^{3}}{r^{2}} \frac{\partial}{\partial r}
\left[\frac{r^{2}\left(\frac{\alpha_{1}}{r} - \alpha_{11}\right)^{\frac{1}{2}}}{(\alpha + \beta)^
{\frac{3}{2}}}\right].
\end{equation}
The non-vanishing component of the acceleration vector 
\begin{equation}
\label{eq35}
\dot{v}_{1} = v_{i;j} v^{j},
\end{equation}
is given by
\begin{equation}
\label{eq36}
\dot{v}_{1} = - \frac{\alpha_{1}}{(\alpha + \beta)}.
\end{equation}
Thus the acceleration is always directed in radial direction and the fluid
flow in $t-$direction is uniform. If $\alpha_{1}< 0$, acceleration is positive
and if $\alpha_{1} > 0$, there will be deceleration.\\
The expression for kinematical parameter expansion $\theta$ is given by
\begin{equation}
\label{eq37}
\theta = 3\beta_{4},
\end{equation}
All components of rotation $w_{ij}$ and shear $\sigma_{ij}$ tensors are found 
to be zero. We observe that  the expansion is time-dependent only. Hence the 
model (20) representing a distribution of charged perfect fluid is expanding 
with time but non-rotating and non-shearing.\\
\section{Bulk viscous models}
In this section bulk viscous models of the universe are discussed. Weinberg
[16] has suggested that in order to consider the effect of bulk viscosity,
the perfect fluid pressure should be replaced by effective pressure $\bar{p}$ by
\begin{equation}
\label{eq38}
\bar{p} = p - \xi \theta,         
\end{equation}
where $p$ represent equilibrium pressure, $\xi$ is the coefficient of bulk viscosity
and $\theta$ is the expansion scalar. Here $\xi$ is, in general, a function of time.
Therefore, from eq. (\ref{eq21}), we obtain
\[
8\pi (p - \xi \theta) - \Lambda = 3(\alpha^{2}_{1} - \beta^{2}_{4}) \] 
\begin{equation}
\label{eq39}
+ (\alpha + \beta)
\left(2\beta_{44} - \alpha_{11} - \frac{3\alpha_{1}}{r}\right).
\end{equation}
Thus, given $\xi(t)$ we can solve the cosmological parameters. In most of the 
investigations involving bulk viscosity is assumed to be a simple power function of 
the energy density [21]-[23]. 
\begin{equation}
\label{eq40}
\xi(t) = \xi_{0} \epsilon^{n},
\end{equation}
where $\xi_{0}$ and $n$ are constants. If $n = 1$, eq. (\ref{eq40}) may correspond
to a radiative fluid. However, more realistic models [24] are based on lying 
in the regime $0 \leq n \leq \frac{1}{2}$.
\subsection {$Model I:  (\xi = \xi_{0})$}
In this case we assume $n = 0$ in eq. (\ref{eq40}) which gives $\xi = \xi_{0}$ = constant.
By the use of eqs. (\ref{eq23}) and (\ref{eq37}) in eqs. (\ref{eq22}) and (\ref{eq39}),
we obtain
\begin{equation}
\label{eq41}
4\pi(1 + \gamma)\epsilon = 12\pi \xi_{0}\beta_{4} + (\alpha + \beta)(\alpha_{11} + \beta_{44})
\end{equation}
\[
(1 + \gamma)\Lambda = 3(1 + \gamma)(\beta^{2}_{4} - \alpha^{2}_{1}) \] 
\[
+ (\alpha + \beta)
\left[(1 + 3\gamma)\alpha_{11} + 3(1 + \gamma)\frac{\alpha_{1}}{r} - 2\beta_{44}\right]\]
\begin{equation}
\label{eq42}
- 24\pi \xi_{0} \beta_{4}
\end{equation}
\subsection {$Model II:  (\xi = \xi_{0}\epsilon)$}
In this case we assume $n = 1$ in eq. (\ref{eq40}) which gives $\xi = \xi_{0}\epsilon$.
By the use of eqs. (\ref{eq23}) and (\ref{eq37}) in eqs. (\ref{eq22}) and (\ref{eq39}),
we obtain
\begin{equation}
\label{eq43}
4\pi \epsilon = \frac{(\alpha + \beta) (\alpha_{11} + \beta_{44})}{(1 + \gamma - 3\xi_{0}\beta_{4})}
\end{equation}
\[
(1 + \gamma)\Lambda = 3(1 + \gamma)(\beta^{2}_{4} - \alpha^{2}_{1}) \] 
\[
+ (\alpha + \beta)
\left[(1 + 3\gamma)\alpha_{11} - 2\beta_{44} + 3(! + \gamma)\frac{\alpha_{1}}{r}\right] \]
\begin{equation}
\label{eq44}
- \frac{6\xi_{0}\beta_{44}(\alpha + \beta) (\alpha_{11} 
+ \beta_{44})}{(1 + \gamma -3\xi_{0}\beta_{4})}
\end{equation}
These eqs. (\ref{eq41}) - (\ref{eq44}) will supply different viable models for
suitable choices of $\beta(t)$.
\section{Conclusions} 
 We have obtained conformally flat spherically symmetric cosmological models 
in the presence of a charged perfect fluid where the acceleration is always 
directed in radial direction and the fluid flow in $t$-direction is uniform. 
We have also discussed two particular cases. In both cases we observe that 
the energy conditions hold good and the cosmological constant is found positive
and is decreasing function of time which is consistent with the present observations.
The model is expanding with time but non-rotating and non-shearing.\\
Assuming $\xi(t) = \xi_{0} \epsilon^{n}$, where $\epsilon$ is the energy density 
and $n$ is the positive index, we have obtained exact solutions. The effect of the 
bulk viscosity is to produce a change in the perfect fluid. 
\section*{Acknowledgements} 
One of the authors (A. Pradhan) would like to thank the Inter-University Centre 
for Astronomy and Astrophysics, India for providing  facility where this work was 
carried out. We would like to thank R. G. Vishwakarma for many helpful discussions. \\
\newline
\newline


\noindent
\begin{thebibliography}{99}
\bibitem {ref1}  H. A. Buchdahl, {\it Phys. Rev.} {\bf 115}, 1325  (1959). 
\bibitem {ref2}  K. P. Singh and S. R. Roy, {\it Proc. Natn. Inst. Sci. India}, {\bf 32A}, 223 (1966).  
\bibitem {ref3}  K. P. Singh and Abdussattar, {\it Gen. Rel. Grav.} {\bf 5}, 115 (1974). 
\bibitem {ref4}  S. R. Roy and Raj Bali, {\it Indian J. pure appl. Math.} {\bf 9}, 871 (1978).
\bibitem {ref5}  S. N. Pandey and R. Tiwari, {\it Indian J. pure appl. Math.} {\bf 12}, 261 (1981). 
\bibitem {ref6}  D. R. K. Reddy, {\it J. Math. Phys.} {\bf 20}, 23 (1979).
\bibitem {ref7}  V. U. M. Rao and D. R. K. Reddy, {\it Gen. Rel. Grav.} {\bf 14}, 1017 (1982). 
\bibitem {ref8}  K. Shanthi, {\it Astrophys. Space Sc.} {\bf 162}, 163 (1989).
\bibitem {ref9}  A. Melfo and H. Rago, {\it Astrophys. Space Sc.} {\bf 193}, 9 (1992).  
\bibitem {ref10}  Philip D. Mannheim, {\it The Astrophys. J.} {\bf 391}, 429 (1992).
\bibitem {ref11}  R. B. S. Yadav and U. Prasad, {\it Astrophys. Space Sc.} {\bf 203}, 37 (1993).   
\bibitem {ref12}  G. Endean,  {\it Astrophys. J.} {\bf 479}, 40 (1997).
\bibitem {ref13}  G. Endean,  {\it J. Math. Phys.} {\bf 39}, 1551 (1998). 
\bibitem {ref14}  V. V. Obukhov, S. D. Odintsov and L. N. Granda,  {\it Europhys. Lett.} 
{\bf 46}, 268 (1999).
\bibitem {ref15}  M. K. Mak and T. Harko,  {\it Int. J. Mod. Phys.} {\bf D9}, 475 (2000).
\bibitem {ref16}  S. Weinberg, ``Gravitation and Cosmology'', J.Wiley and Sons, New York, 1972.
\bibitem {ref17}  W. Israel and J. N. Vardalas, {\it Lett. nuovo Cim.} {\bf 4}, 887 (1970).
\bibitem {ref18}  Z. Klimek, {\it Post. Astron.} {\bf 19}, 165 (1971).
\bibitem {ref19}  S. Weinberge, {\it Astrophys. J.} {\bf 168}, 175 (1971).
\bibitem {ref20}  L. Landau and E. M. Lifshitz, ``Fluid Mechanics'', Addison-Wisley, 
Mass. 1962, p. 304.
\bibitem {ref21}  D. Pavon, J. Bafaluy and D. Jou, {\it Class Quantum Gravit.} {\bf 8}, 357 (1991);
``Proc. Hanno Rund Conf. on Relativity and Thermodynamics'', Ed. S. D. Maharaj, University of Natal,
Durban, (1996), p. 21.
\bibitem {ref22}  R. Maartens, {\it Class Quantum Gravit.} {\bf 12}, 1455 (1995). 
\bibitem {ref23}  W. Zimdahl, {\it Phys. Rev.} {\bf D53}, 5483 (1996).
\bibitem {ref24}  N. O. Santos, R. S. Dias and A. Banerjee, {\it J. Math. Phys.} {\bf 26}, 878 (1985).
\bibitem {ref25}  T. Padmanabhan and S. M. Chitre, {\it Phys. Lett.} {\bf A120}, 433 (1987).
\bibitem {ref26}  V. B. Johri and R. Sudarshan, {\it Phys. Lett.} {\bf A132}, 316 (1988).
\bibitem {ref27}  A. Pradhan, R. V. Sarayakar and A. Beesham, {\it Astr. Lett. Commun.} 
{\bf 35}, 283 (1997).
\bibitem {ref28}  A. Pradhan, V. K. Yadav and I. Chakrabarty, {\it Int. J. Mod. Phys.} {\bf D10},
339 (2001); {\it Int. J. Mod. Phys.} {\bf D11}, 857 (2002).
\bibitem {ref29}  I. Chakrabarty, A. Pradhan and N. N. Saste, {\it Int. J. Mod. Phys.} {\bf D10},
741 (2001).
\bibitem {ref30}  A. Pradhan and V. K. Yadav,  {\it Int. J. Mod. Phys.} {\bf D11}, 857 (2002).
\bibitem {ref31}  A. Pradhan and Aotemshi I.,  {\it Int. J. Mod. Phys.} {\bf D11}, (2002), in press.
\bibitem {ref32}  \O. Gr\o n, {\it Astrophys. Space Sc.} {\bf 173}, 191 (1990). 
\bibitem {ref33}  S. Perlmutter {\it et al.}, {\it Astrophys. J.} {\bf 483}, 565 (1997), Supernova Cosmology 
Project Collaboration (astro-ph/9608192); {\it Nature} {\bf 391}, 51 (1998), Supernova Cosmology 
Project Collaboration (astro-ph/9712212); {\it Astrophys. J.} {\bf 517}, 565 (1999), 
Project Collaboration (astro-ph/9608192).
\bibitem {ref34} A. G. Riess {\it et al.}, {\it Astron. J.} {\bf 116}, 1009 (1998); 
Hi-Z Supernova Team Collaboration (astro-ph/9805201).
\bibitem {ref35} P. M. Garnavich {\it et al.}, {\it Astrophys. J.} {\bf 493}, L53 (1998a),
Hi-Z Supernova Team Collaboration (astro-ph/9710123); {\it Astrophys. J.} {\bf 509}, 74 (1998b);
Hi-Z Supernova Team Collaboration (astro-ph/9806396).
\bibitem {ref36} B. P. Schmidt {\it et al.}, {\it Astrophys. J.} {\bf 507}, 46 (1998),
Hi-Z Supernova Team Collaboration (astro-ph/9805200). 
\end{thebibliography}
\end{document}